\documentclass[%
 reprint,
nofootinbib,
 amsmath,amssymb,
 aps,
 prd
]{revtex4-2}

\usepackage{graphicx}
\usepackage{dcolumn}
\usepackage{bm}
\usepackage{hyperref}
\usepackage{physics}
\usepackage{multirow}
\usepackage{subcaption}
\usepackage{braket}
\usepackage{slashed}
\usepackage{xcolor}
\usepackage[normalem]{ulem}

\newcommand{\sigmav}{\langle\sigma v\rangle}

\def\L{\mathcal{L}}
\def\MPL{M_{\rm Pl}}
\def\bPsi{\bar\Psi}

\def\bZ{\mathbb{Z}}
\def\hc{\text{h.c.}}
\def\vew{v_{\rm EW}}
\def\xfo{x_{\rm FO}}
\def\Tfo{T_{\rm FO}}
\def\Trh{T_{\rm RH}}
\renewcommand{\Tr}{\mathrm{Tr}}

\begin{document}

\title{A Phase-Transition Portal to Enhanced Dark Matter Annihilation}

\author{Peisi Huang}
\email{peisi.huang@unl.edu}
\affiliation{Department of Physics and Astronomy, University of Nebraska, Lincoln, NE 68588, USA}

\author{Anibal D. Medina}
\email{anibal.medina@fisica.unlp.edu.ar}
\affiliation{IFLP, CONICET - Dpto. de F\'{\i}sica, Universidad Nacional de La Plata, C.C. 67, 1900 La Plata, Argentina}

\author{Carlos E. M. Wagner}
\email{cwagner@uchicago.edu}
\affiliation{HEP Division, Argonne National Laboratory, 9700 Cass Ave., Argonne, IL 60439, USA}
\affiliation{Enrico Fermi Institute, Physics Department, University of Chicago, Chicago, IL 60637, USA}
\affiliation{Kavli Institute for Cosmological Physics, University of Chicago, Chicago, IL 60637, USA}
\affiliation{Leinweber Institute for Theoretical Physics, University of Chicago, Chicago, IL 60637, USA}
\affiliation{Perimeter Institute for Theoretical Physics, Waterloo, ON N2L 2Y5, Canada}

\author{Kairui Zhang}
\email{kzhang25@ou.edu}
\affiliation{Homer L. Dodge Department of Physics and Astronomy, University of Oklahoma, Norman, OK 73019, USA}

\begin{abstract}

    We present a thermal dark matter (DM) scenario in which a phase transition after freeze-out changes both the DM spectrum and the dominant annihilation process, allowing an enhanced present day annihilation rate with suppressed direct detection. We realize this mechanism in a two-Higgs-doublet model extended by a complex scalar singlet, a Majorana fermion, and a heavier Dirac fermion. Before the transition, the Majorana state freezes out through $p$-wave annihilation into singlet scalars. The subsequent transition changes the dark fermion masses and mixing, enabling $s$-wave annihilation through a pseudoscalar portal. The surviving relic differs in mass and composition from the original state. We find relic compatible solutions with present day annihilation rates more than an order of magnitude above the canonical thermal value, without tuning to a mediator resonance, while satisfying direct detection bounds.

\end{abstract}

\maketitle

\section{Introduction}

The nature of dark matter~(DM) remains one of the central open questions in particle physics. In the conventional weakly interacting massive particle~(WIMP) picture, chemical freeze-out~(FO) relates the relic density to the annihilation rate. To match with the observed relic density, a canonical value is~\cite{Kolb:1990vq,Steigman:2012nb}
\begin{equation}\label{eq:sigmav_fo}
    \sigmav_{\rm FO} \simeq (2\text{-}3)\times10^{-26}~{\rm cm^3\, s^{-1}}.
\end{equation}
This, however, does not imply that the DM annihilation cross section in the present Universe should be in this range. Quite often, a smaller annihilation cross section today arises readily in conventional $p$-wave scenarios, owing to the lower DM velocities. Here instead, we investigate the reverse hierarchy: a thermal relic whose present day annihilation cross section is substantially larger than at FO, but without a correspondingly large signal in direct detection~(DD). This is accomplished in our scenario by a phase transition that changes the dominant annihilation from $p$-wave to $s$-wave. In fact, neither the DM mass spectrum nor the dominant annihilation process need remain unchanged between FO and the present Universe. The phase transition also changes the DM spectrum, allowing the relic today to differ in mass and composition from the FO state, see for example ~\cite{Huang:2022vkf,Huang:2026qdc}.

In order to realize this mechanism, we study a fermionic dark sector coupled to a complex scalar $\phi$, singlet under the SM gauge groups and with a portal interaction with an enlarged Higgs sector. The singlet acquires a vacuum expectation value~(VEV) through a delayed phase transition after FO. Before the transition, DM freezes out in the portal's symmetric phase through $p$-wave annihilation into singlet scalars. The transition later changes the dark fermion masses and mixing, enabling the surviving DM state to couple to the singlet's CP-odd component. Through a pseudoscalar portal to Standard Model~(SM) fermions, this coupling opens an $s$-wave annihilation channel that remains unsuppressed at low velocities. The same CP-odd interaction produces momentum suppressed spin-dependent elastic scattering on nuclei at tree level~\cite{Abe:2018emu}, allowing a large present day annihilation rate with a small contribution to the DD signal.

Connecting a pseudoscalar to the SM requires additional structure. SM gauge invariance forbids its direct renormalizable couplings to SM fermions or gauge boson pairs, while CP conservation forbids its mixing with the SM Higgs boson. The portal can therefore only be realized through an extended Higgs sector or through higher dimensional interactions of the type familiar from axion-like particles~(ALPs)~\cite{Bauer:2020jbp}. A two-Higgs-doublet model~(2HDM) provides a physical CP-odd scalar with Yukawa couplings to SM fermions, which the gauge-singlet pseudoscalar inherits through mixing, while an ALP can couple to SM particles through interactions generated by integrating out heavy particles charged under the SM gauge group. In this paper, we choose to extend the SM Higgs sector to a type-II 2HDM as a concrete implementation. The phase transition mechanism and its qualitative conclusions can easily extend to an ALP-like portal with the same dark sector evolution and relevant pseudoscalar interactions.

Recently, Ref.~\cite{Totani:2025fxx} reported a Galactic halo excess~(GHE) in an analysis of 15 years of \textit{Fermi}-LAT data, with a gamma ray spectrum peaking near $20~{\rm GeV}$. The analysis focuses on regions away from the Galactic disk and center, and the reported emission is distinct from the lower energy Galactic Center excess~(GCE)~\cite{Goodenough:2009gk,Hooper:2010mq}. Under a DM interpretation, representative fits favor masses of $0.3$--$0.8~{\rm TeV}$ and an annihilation cross section in the Milky Way~(MW) halo into $b\bar b$ or $W^+W^-$ of order
\begin{equation}
    \sigmav_{\rm MW}\sim(5\text{-}8)\times10^{-25}~{\rm cm^3\, s^{-1}},
\end{equation}
with the inferred values subject to halo modeling~\cite{Totani:2025fxx,Stenhouse:2026nve}. Although no known astrophysical objects can mimic such a hard peak, the DM interpretation faces several tensions. In particular, $\sigmav_{\rm MW}$ is more than an order of magnitude above the canonical FO rate $\sigmav_{\rm FO}$ in Eq.~\eqref{eq:sigmav_fo}. As an application, our model, which predicts an enhanced present day annihilation rate and suppressed nuclear scattering, is well suited to address the tensions plaguing the DM interpretation of the GHE.

A DM interpretation of GHE must reconcile four quantities: the canonical freeze-out rate $\sigmav_{\rm FO}$, the rate $\sigmav_{\rm MW}$ inferred from the halo signal, the rate $\sigmav_{\rm dSph}$ constrained by dwarf spheroidal galaxies, and the elastic-scattering cross section constrained by DD. Besides the tension between $\sigmav_{\rm MW}$ and $\sigmav_{\rm FO}$, a comparable rate in dwarfs can be in conflict with their gamma-ray limits~\cite{Circiello:2026inp}, subject to the uncertainties in their DM distributions. Interactions invoked to explain the GHE must also respect stringent DD limits~\cite{LZ:2024zvo,XENON:2025vwd}. Resolving one tension does not establish consistency with the other two.

The usual ways of evading DD bounds do not work here. For example, for a subcomponent tracing the halo with fraction $f$, DD bounds relax as $f^{-1}$, while the GHE requires $\sigmav_{\rm MW}\propto f^{-2}$, worsening the tension. Crossing symmetry also correlates DD with the portal interactions responsible for annihilation. For a standard minimal portal generated through mixing with the Higgs, a cancellation in the tree-level scattering amplitude with SM particles usually implies a suppressed annihilation amplitude by crossing symmetry, unless one of the processes lies at a kinematic threshold~\cite{Cabrera:2019gaq}.

Resonant annihilation can enhance annihilation without comparably enhancing elastic scattering. Reconciling the different annihilation rates with DD bounds nevertheless requires examining the interactions responsible for each stage of the thermal history. For example, Ref.~\cite{Murayama:2025res} uses a narrow resonance to reconcile the annihilation rates in the MW halo and dwarf galaxies, while a separate Higgs portal interaction sets the thermal abundance. The associated Higgs-mediated spin-independent scattering introduces a potential tension with DD bounds. Ref.~\cite{Nomura:2026ntp} addresses the tension among $\sigmav_{\rm MW}$, $\sigmav_{\rm dSph}$, and DD, with an additional astrophysical boost required to reconcile its thermal relic abundance with the GHE normalization.

On the other hand, a few $p$-wave Sommerfeld models proposed to explain these tensions~\cite{Jho:2025iah,Yoshimatsu:2026gff} fail to establish relic density consistency because their FO estimates do not consistently track kinetic decoupling and dangerous reannihilation. As has long been recognized, kinetic decoupling accelerates DM cooling in scenarios with Sommerfeld enhancement and, near a Sommerfeld resonance, can restore efficient annihilation long after FO, further depleting the relic and making the final abundance highly sensitive to the kinetic history~\cite{Feng:2010zp,Binder:2018znk,Binder:2017lkj,Binder:2017rgn,vandenAarssen:2012ag,Binder:2021bmg}. Therefore, these $p$-wave Sommerfeld-type models require careful Boltzmann evolution with velocity dependent distributions. Naive canonical FO-rate matching and unjustified kinetic equilibrium are untenable. Neglecting kinetic decoupling and reannihilation can lead to orders of magnitude overestimates of the benchmark relic abundances~\cite{Jho:2025iah,Yoshimatsu:2026gff}.

We apply our model to the GHE as an illustrative example. The delayed phase transition provides a dynamical separation between the interactions setting the thermal abundance and those governing present day annihilation. We study how the accompanying changes in the DM spectrum and partial wave structure can produce an enhanced halo signal while keeping the contribution to DD very suppressed.

The rest of this paper is organized as follows. In Sec.~\ref{sec:model}, we present the model structure. Sec.~\ref{sec:sym_phase_FO} discusses the symmetric phase thermal history and the FO calculation, and Sec.~\ref{sec:broken_phase_halo} the broken phase spectrum and the resulting halo annihilation. The phenomenological results are presented in Sec.~\ref{sec:results}. In Sec.~\ref{sec:discussion} we discuss direct detection and dwarf spheroidal constraints, and we conclude in Sec.~\ref{sec:conclusions}.

\section{Model}\label{sec:model}

\begin{table}[htb!] 
    \centering 
    \large 
    \begin{tabular}{p{1.5cm} p{0.75cm} p{0.75cm} p{0.75cm} p{0.75cm} p{0.75cm} p{0.75cm} p{0.75cm}} 
        \hline 
         & $\phi$ & $H_{1}$ & $H_2$ & $\chi_M$ & $\chi_L$ & $\chi_R$ & $u_R$ \\ 
        \hline 
        \hline 
        $\quad \bZ_3$ & 2 & 0 & 1 & 0 & 1 & 1 & 1  \\ 
        \hline 
        $\quad \bZ_2$ & + & + & + & + & + & + & - \\
        \hline 
    \end{tabular} 

    \caption{The $\bZ_3$ assignments result in a type-II 2HDM Yukawa structure for the SM fermions. All SM fermions are odd under the $\bZ_2$ parity and all other particles are even.}
    \label{tab:charges} 
\end{table}

We consider a type-II 2HDM~\cite{Branco:2011iw} extended by a complex scalar singlet $\phi$, a Majorana fermion $\chi_M$, and two Weyl fermions $\chi_{L,R}$ forming a Dirac fermion $\chi_D$\footnote{As it happens, the field contents resemble a partially decoupled next-to-minimal supersymmetric Standard Model~(NMSSM) spectrum~\cite{Carena:2015moc,Slavich:2020zjv,Baer:2024hpl,Baer:2024fgd,Baer:2026ytu,Baer:2024tfo,Baer:2023olq,Baer:2025zqt,Maniatis:2009re,Ellwanger:2009dp}.}. The charge assignments are shown in Table~\ref{tab:charges}.

The $s$-wave initial state of two identical Majorana fermions has $J=0$ and is CP odd. A $\phi\phi^*$ final state with $J=0$ is CP even, so CP conservation forbids this $s$-wave transition, and the leading allowed annihilation is $p$-wave. We impose $\mathbb{Z}_3$ which allows $\chi_M\chi_M\to\phi\phi^*$ while forbidding the same charge final states $\phi\phi$ and $\phi^*\phi^*$. These same charge states are exchanged by CP, and their linear combinations include a CP-odd state that could otherwise allow $s$-wave annihilation. The $\mathbb{Z}_3$ is the smallest cyclic symmetry that distinguishes these channels. The same charge assignment gives the type-II 2HDM Yukawa structure and forbids tree-level flavor changing neutral currents~\cite{Glashow:1976nt,Paschos:1976ay}. A mediator carrying nonzero $\mathbb{Z}_3$ charge cannot have a symmetry preserving Majorana mass; we therefore introduce two Weyl fermions that form a Dirac mediator.

The $\bZ_2$ ensures the DM's stability: only SM fermions are odd, while all other fields are even. If the lightest dark fermion were to decay, the process always involves an odd number of SM fermions, which would violate the $\bZ_2$\footnote{The $\bZ_2$ parity has a well motivated UV origin, which can be a discrete leftover of a gauged $U(1)_{B-L}$ and forbids even higher-dimension decay operators~\cite{Sheng:2026aro}. Other UV origins such as an accidental leftover of a supersymmetric discrete $R$-symmetry are also possible~\cite{Baer:2025srs,Baer:2025oid}}.

\subsection{Scalar sector}

The scalar potential relevant to the DM analysis is
\begin{equation}\label{eq:V_scalar}
    V \supset V_\phi + V_{\rm 2HDM} + V_{H\phi},
\end{equation}
where
\begin{equation}\label{eq:V_phi}
\begin{split}
    V_\phi
        &\supset m_\phi^2|\phi|^2 + \frac{\lambda_\phi}{4}|\phi|^4 + \left( \frac{\mu_\phi}{3!}\phi^3 + \hc \right),
\end{split}
\end{equation}
\begin{equation}
\begin{split}
    V_\text{2HDM} 
        &= m_{11}^2 H_1^\dagger H_1 + m_{22}^2 H_2^\dagger H_2 - (m_{12}^2H_1^\dagger H_2 + \hc) \\ 
            &\qquad + \frac{\lambda_1}{2}(H_1^\dagger H_1)^2 + \frac{\lambda_2}{2}(H_2^\dagger H_2)^2 \\ 
                &\qquad\quad + \lambda_3(H_1^\dagger H_1)(H_2^\dagger H_2) + \lambda_4(H_1^\dagger H_2)(H_2^\dagger H_1),
\end{split}
\label{eq:V_2hdm}
\end{equation}
and\footnote{Certain $\bZ_3$-compatible quartic interactions such as $\phi^2H_2^\dagger H_1$ are dropped for reasons explained in the next section.}
\begin{equation}
\begin{split}
    V_{H\phi} 
        &= \sum_{i=1,2}\lambda_{\phi i}|\phi|^2(H_i^\dagger H_i) + \left(\kappa_{12}\phi H_1^\dagger H_2 + \hc \right).
\end{split}
\label{eq:V_hphi}
\end{equation}
Eqs.~\eqref{eq:V_phi}--\eqref{eq:V_hphi} give the zero-temperature terms relevant to our DM analysis. Additional contributions needed for the phase transition are left unspecified and are assumed not to modify the zero-temperature spectrum and mixing appreciably. The $\kappa_{12}$ interaction mixes the CP-odd states $a_0$ and $A_0$, generating the pseudoscalar portal for present day halo annihilation. After the singlet transition,
\begin{equation}
    \phi = \frac{1}{\sqrt2}(v_\phi+s_0+i a_0).
    \label{eq:phi_expansion}
\end{equation}
The broken-phase scalar spectrum contains three CP-even states $h,H,s$, and two CP-odd states $A,a$. The detailed scalar sector mixing calculations are given in Appendix~\ref{app:scalar_sector}.

\subsection{Fermion sector}

The dark fermion Lagrangian is
\begin{equation}\label{eq:fermion_lagrangian}
\begin{split}
    \L_{\chi} 
        &\supset -\frac{M_\chi}{2}\chi_M\chi_M - M_D\chi_L\chi_R^c - y_L \phi\chi_M\chi_L - y_R\phi^*\chi_M\chi_R^c \\ 
            &\qquad\quad - \frac{y_{LL}}{2}\phi^*\chi_L\chi_L - \frac{y_{RR}}{2}\phi\chi_R^c\chi_R^c + \hc \\
        &= - \frac{M_\chi}{2}\bPsi_M\Psi_M - M_D\bPsi_D\Psi_D - \phi\bPsi_M\Gamma\Psi_D - \phi^*\bPsi_D\Gamma'\Psi_M \\ 
            &\qquad\quad -\frac{1}2{} \left(y_{LL}\phi^*\overline{\Psi^c_D}P_L\Psi_D + y_{RR} \phi \bar\Psi_D P_L\Psi_D^c + \hc \right),
\end{split}
\end{equation}
where
\[
    \Gamma \equiv y_LP_L + y_R P_R, \quad \Gamma' \equiv \gamma^0\Gamma\gamma^0 = y_R P_L + y_L P_R,
\]
and we also embed the two-component Weyl fermions into four-component notation $\Psi$, where  $\Psi_D \equiv \begin{pmatrix}
    \chi_L \\
    (\chi_R^{c})^\dagger
\end{pmatrix}$ is a Dirac fermion mediating the annihilation and $\Psi_M$ is the four-components Majorana spinor DM candidate, built from $\chi_M$. CP symmetry is imposed to ensure all the couplings are real. The charge assignments in Table~\ref{tab:charges} forbid a diagonal Yukawa coupling between $\phi$ and $\chi_M$. 

\section{Symmetric Phase and Freeze-out}\label{sec:sym_phase_FO}

\subsection{Singlet Phase Transition}\label{subsec:phi_PT}

After the singlet $\phi$'s phase transition, the dominant annihilation changes from $p$-wave to $s$-wave. Thus, it's necessary to assume that the transition temperature $T_*$ and more importantly, the reheating temperature $\Trh$ to be sufficiently below $\Tfo$ such that the newly opened channel does not appreciably re-deplete the frozen abundance. This requires~\cite{Cohen:2008nb}
\begin{equation}
    \frac{\Trh}{\Tfo} \ll \frac{\sigmav_{\rm FO}}{2\sigmav_0}.
    \label{eq:reannhi}
\end{equation}
The decrease in the relativistic degrees of freedom between FO and reheating further weakens this condition. The GHE-motivated parameter region in Ref.~\cite{Totani:2025fxx} is best fitted by DM annihilation in the mass range 
\[
    m_\chi \sim 0.3\text{--}0.8~{\rm TeV}.
\]
This generally requires the reheating temperature $\Trh \lesssim \order{1}$ GeV for negligible reannihilation. Because the singlet $\phi$ mixes with the EW Higgs sector, the singlet VEV $v_\phi$ is larger than this $\Trh$. This hierarchy indicates that the transition might be supercooled. A supercooling can arise through different mechanisms, including tree-level potential barriers generated by cubic scalar interactions~\cite{Athron:2022mmm}, or nearly scale invariant potentials with radiative Coleman–Weinberg corrections~\cite{Marzola:2017jzl,Jinno:2016knw}. The energy released by the transition can also reheat the plasma and dilute the frozen DM abundance through entropy production. 
\begin{equation}
    T_{\rm RH}\simeq T_*(1+\alpha)^{1/4}, 
    \quad 
    \Delta=(1+\alpha)^{3/4}.
\end{equation}
Here $\alpha$ is the strength of the phase transition, and $\Delta$ is the dilution factor. In the following discussion, we consider both the negligible dilution limit, $\alpha\ll1$, and a benchmark with substantial entropy production, $\alpha = 10$. They correspond to dilution factors $\Delta\simeq1$ and $\Delta\simeq 6$, respectively. Note however that the hierarchy $v_\phi \gg$ $T_*$ does not always imply strong supercooling~\cite{Cohen:2008nb,Heurtier:2019beu}. In addition, the hierarchy $v_\phi\gg\Trh$ is tied to the GHE-motivated parameter region and a 2HDM realization. Here, the large $v_\phi$ arises from the singlet's mixing with the electroweak scalar sector. Other realizations, such as an ALP-like portal, can provide the required pseudoscalar interactions without this mixing and hence, no similarly large VEV. We leave such realizations for future work. Heavier freeze-out DM can allow higher transition temperatures, also avoiding the need for supercooling. Supercooling is not intrinsic to the $p$- to $s$-wave mechanism.

Since the singlet phase transition occurs below electroweak symmetry breaking~(EWSB), one might worry that the cubic interaction $(\kappa_{12}\phi H_1^\dagger H_2+\hc)$ induces an effective tadpole for $\phi$ and triggers the singlet transition prematurely. This tadpole can be avoided by embedding the $\bZ_3$ into a larger $\bZ_6$ symmetry, under which the bare cubic interaction is forbidden while $S\phi H_1^\dagger H_2$ is allowed. The cubic interaction is then generated only after a hidden singlet $S$ acquires a VEV below the EWSB scale and after FO~\footnote{ The hidden singlet may not be a fundamental field, but a condensate of a hidden confinement sector with a large number of dark flavors~\cite{Agrawal:2025wvf,Konstandin:2011dr}. Additional dark fermions coupled to $\phi$ provide thermal corrections that compensate the negative bare mass squared $m_\phi^2$. A large number of flavors can also maintain this thermal trapping down to lower temperatures~\cite{Cohen:2008nb}.}. For example, we can assign
\[
    (q_S, q_\phi, q_{H_1}, q_{H_2}, q_{\chi_M}, q_{\chi_L}, q_{\chi_R}, q_{u_R})
        = (3, 2, 0, 1, 0, 4, 4, 1)
\]
under $\bZ_6$. These charges reduce modulo $3$ to the $\bZ_3$ assignments in Table~\ref{tab:charges}.

The same $\bZ_6$ forbids the bare quartic portal $\phi^2H_2^\dagger H_1$, which is why we omit it from the scalar potential Eq.~\eqref{eq:V_scalar}. All the remaining terms respect this $\bZ_6$, except for $m_{12}^2$, which we still retain as an explicit soft breaking term to lift the vacuum degeneracy and avoid potential stable domain walls.

\subsection{Freeze-out Abundance}

\begin{figure}[t]
    \centering
    \includegraphics[width=0.6\linewidth]{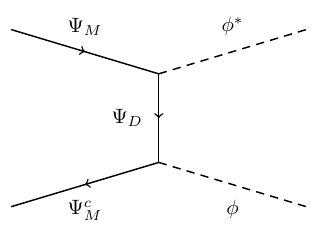}
    \caption{The $t$-channel number changing diagram of $\Psi_M$ during freeze-out.
     }
    \label{fig:chi_t_diagrams}
\end{figure}

During FO, the singlet $\phi$ has not yet acquired a VEV, and the dark fermions remain unmixed. The $\bZ_3$ charge assignment forbids a diagonal Yukawa coupling between $\phi$ and $\chi_M$, so the dominant FO process is $\chi_M\chi_M\to\phi\phi^*$ through $t$- and $u$-channel exchange of the Dirac $\chi_D$. This annihilation is $p$-wave. A representative diagram is shown in Fig.~\ref{fig:chi_t_diagrams}.

We assume that the thermal scalar mass during FO satisfies $m_\phi(\Tfo)<M_\chi$, so that the annihilation channel is kinematically open. We neglect the thermal scalar masses in the approximate FO calculation. Phase-space corrections become important only if the thermal scalar masses happen to be near the annihilation threshold. As shown in Sec.~\ref{sec:results}, $M_\chi$ can reach a few TeV and can be substantially heavier than $|m_1|$. The explicit cross sections are given in Appendix~\ref{app:ann_xsec}.

To suppress possible $s$-wave coannihilation during FO, we also impose
\begin{equation}
    M_D-M_\chi>5\,\Tfo.
\end{equation}
The Dirac $\chi_D$ abundance is thus Boltzmann suppressed during $\chi_M$'s FO~\cite{Griest:1990kh}. We can safely neglect coannihilation and determine the frozen abundance from $\chi_M$ annihilation.

The asymptotic comoving abundance is~\cite{Kolb:1990vq,Griest:1990kh}
\begin{equation}
    Y^\infty_{\chi_M}
        \simeq
        \frac{\sqrt{45}}
        {\sqrt{\pi g_*(\Tfo)}\,\MPL M_\chi J(\xfo)},
    \label{eq:number_density}
\end{equation}
with
\begin{equation}
    J(\xfo)
        =
        \int_{\xfo}^{\infty}
        \frac{\braket{\sigma v_{\rm rel}}}{x^2}\,dx
        = \frac{\braket{\sigma v_{\rm rel}}_{\rm FO}}{2\xfo}.
\end{equation}
The FO parameter can be obtained iteratively as~\cite{Kolb:1990vq}
\begin{equation}
    \xfo
        =
        \ln\left[
        \frac{
            0.038\,g_\chi\MPL M_\chi\braket{\sigma_{\rm eff}v}_{\rm FO}
        }{
            g_*^{1/2}(\Tfo)\xfo^{1/2}
        }
        \right],
\end{equation}
where $g_\chi=2$ for a Majorana fermion and $g_*$ is the number of relativistic degrees of freedom at decoupling.

\section{Broken Phase}\label{sec:broken_phase_halo}

\subsection{Dark fermion masses and couplings}
After the singlet phase transition, the VEV $v_\phi$ generates additional entries in the dark fermion mass matrix. We define
\[
    m_{L} \equiv \frac{y_L v_\phi}{\sqrt{2}}, \ m_{R} \equiv \frac{y_R v_\phi}{\sqrt{2}}, \ m_{LL}\equiv \frac{y_{LL} v_\phi}{\sqrt{2}}, \ m_{RR} \equiv \frac{y_{RR} v_\phi}{\sqrt{2}}.
\]
In the basis $(\chi_L, \chi_R^c, \chi_M)$, the mass matrix is written as
\begin{equation}
    M=
    \begin{pmatrix}
        m_{LL} & M_D & m_L \\
        M_D & m_{RR} & m_R \\
        m_L & m_R & M_\chi
    \end{pmatrix}.
\end{equation}
There are three mass eigenstates in the broken phase with eigenvalues $m_{1,2,3}$, $|m_1|<|m_2|<|m_3|$. Let
\[
    \vec{\chi}_0 \equiv (\chi_L\ \chi_R^c\ \chi_M)^T, \qquad \vec{\chi} \equiv (\chi_1\ \chi_2\ \chi_3)^T.
\]
Then, the Yukawa parts of Eq.~\eqref{eq:fermion_lagrangian} can be written as
\begin{equation}
    \L_Y 
        = -\frac{1}{2}\vec\chi_0 ^T (\phi Y_\phi + \phi^* Y_{\phi^*})\vec\chi_0 + \hc,
\end{equation}
where
\begin{equation}
     Y_\phi
        =
        \begin{pmatrix}
            & & y_L \\
            & y_{RR} &  \\
            y_L & &
        \end{pmatrix},
    \quad
    Y_{\phi^*}
        =
        \begin{pmatrix}
           y_{LL} & &  \\
            & & y_R \\
            & y_R &
        \end{pmatrix}.
\end{equation}
The mass matrix $M$ is diagonalized by a unitary matrix $U$
\begin{equation}\label{eq:mass_diag}
    \vec\chi_0 = U\vec\chi, \quad \text{diag}(|m_1|, |m_2|, |m_3|) = U^T M U.
\end{equation}
The signs in the mass eigenvalues can be rotated away by $U$ but become physical in the interaction couplings. Then,
\begin{equation}
    \hat Y_\phi = U^T Y_\phi U, \quad \hat Y_{\phi^*} = U^T Y_{\phi^*} U
\end{equation}
Expanding $\phi = \frac{1}{\sqrt{2}}(v_\phi + s_0 + ia_0)$,
\begin{equation}
    Y_{s_0} = \hat Y_\phi + \hat Y_{\phi^*}, \quad Y_{a_0} = \hat Y_\phi - \hat Y_{\phi^*},
\end{equation}
one has
\begin{equation}
    \L_Y 
        \supset -\frac{1}{2}\vec\chi^T \left(\frac{Y_{s_0}}{\sqrt{2}}s_0 + i\frac{Y_{a_0}}{\sqrt{2}}a_0\right)\vec\chi + \hc.
\end{equation}
In the following, we use the shorthands
\begin{equation}
    y_{ij}^{s_0} \equiv (Y_{s_0})_{ij} , \quad y_{ij}^{a_0} \equiv (Y_{a_0})_{ij}.
\end{equation}

\subsection{Heavier state decay and relic density}

After the singlet phase transition, the populations of the broken-phase mass eigenstates depend on the evolution of the dark fermion mixing and the thermal history.

In the idealized sudden conversion limit, the post-transition yield is
\begin{equation}
    Y_{\chi_1}
        =Y^\infty_{\chi_M}
        \sum_{i=1}^3 |U_{3i}|^2\nu_{\chi_i},
    \qquad
    \sum_i |U_{3i}|^2=1,
\end{equation}
where $Y^\infty_{\chi_M}$ is given in Eq.~\eqref{eq:number_density}, and $U$ is the unitary matrix used to diagonalize the mass matrix in Eq.~\eqref{eq:mass_diag}. The $\nu_{\chi_i}\geq1$ denotes the mean number of $\chi_1$ particles produced by the decay chain of a $\chi_i$. For example, $\chi_j\to\chi_1a$ followed by $a\to\chi_1\chi_1$ produces three $\chi_1$ when kinematically allowed. One also needs populated heavier states $\chi_{2,3}$ to decay before big-bang nucleosynthesis~(BBN), and impose~\cite{Kolb:1990vq,Kawasaki:2017bqm}
\begin{equation}\label{eq:BBN_decay}
    \Gamma_{\chi_{2,3}}^{\rm tot}
        \gtrsim H(T_{\rm BBN})
        =1.66\sqrt{g_*(T_{\rm BBN})}
         \frac{T_{\rm BBN}^2}{\MPL},
\end{equation}
where $T_{\rm BBN}\simeq1~{\rm MeV}$, $g_*(T_{\rm BBN})=10.75$, and $\MPL=1.22\times10^{19}~{\rm GeV}$.

On the other hand, if $\chi_M$ evolves continuously into $\chi_1$, as in the adiabatic limit, each frozen $\chi_M$ yields exactly one $\chi_1$ in the end, with no heavier mass eigenstates physically populated. The yield is then simply
\begin{equation}\label{eq:Y_adia}
    Y_{\chi_1}=Y^\infty_{\chi_M}.
\end{equation}

For conservativeness, in the scan presented in Sec.~\ref{sec:results}, we adopt this adiabatic limit Eq.~\eqref{eq:Y_adia}, which gives the minimum $\chi_1$ multiplicity of unity, but still apply Eq.~\eqref{eq:BBN_decay} as an additional conservative requirement. The decay calculations are given in Appendix~\ref{app:chi_decay}. Including potential dilution, the final relic density is~\cite{Baer:2026wre,Sheng:2026tqt}
\begin{equation}
    \Omega_{\chi_1}h^2
        =\frac{s_0}{\rho_c/h^2}|m_1|
         \frac{Y_{\chi_1}}{\Delta}
        =2.742\times10^8
        \left(\frac{|m_1|}{\rm GeV}\right)
        \frac{Y_{\chi_1}}{\Delta},
\end{equation}
where $|m_1|$ is the physical mass of $\chi_1$, $s_0$ and $\rho_c$ are the present entropy density and critical density, respectively, and $\Delta$ is the dilution factor. Mild dilution after the phase transition can be compensated by allowing moderate overproduction during FO.

As discussed in Sec.~\ref{subsec:phi_PT}, entropy production dilutes the relic abundance by $\Delta=(1+\alpha)^{3/4}$. This dilution can be easily compensated by increasing $M_D$, which suppresses annihilation in the symmetric phase and increases the initial yield, while adjusting $y_{LL}$ and $y_{RR}$ to retain a viable broken-phase spectrum. As presented in Appendix~\ref{app:ann_xsec}, in the large $M_D$ limit, the annihilation rate scales as $\sigmav_{\rm FO}\propto M_D^{-2}$. To compensate the dilution factor $\Delta$, one can scale up 
\begin{equation}
    M_D \to M_D(1 + \alpha)^{3/8}.
\end{equation}
As an illustration of large entropy dilution, for $\alpha=10$, increasing $M_D$ by a factor of approximately $2.5$ can compensate for the dilution.

\subsection{Broken phase scalar mixing}

In the broken phase, the pure CP-odd singlet state $a_0$ and the pure CP-odd EW doublet state $A_0$ mix. They are related to the eigenstates after mixing by
\begin{eqnarray}
    a_0 &=& s_{\theta_{A}}A + c_{\theta_A}a,\\
     A_0 &=& c_{\theta_A}A - s_{\theta_A}a.
\end{eqnarray}
where $\theta_A$ is the CP-odd mixing angle.

The type-II 2HDM pseudoscalar couplings to SM fermions are~\cite{Branco:2011iw}
\begin{equation}
    \L_{A_0ff}
        =
        \frac{iA_0}{\vew}
        (
            m_d\tan\beta\,\bar d\gamma_5 d
            + m_\ell\tan\beta\,\bar\ell\gamma_5\ell
            + m_u\cot\beta\,\bar u\gamma_5u
        ),
\end{equation}
where $\tan\beta \equiv v_2/v_1$. For $b\bar b$ final states to explain the GHE, after mixing, the relevant terms become
\begin{equation}
    \L_{b\bar b}
        = i\frac{m_b\tan\beta}{\vew}
        (c_{\theta_A}A-s_{\theta_A}a)\bar b\gamma_5 b.
\end{equation}
Similarly, embedding the lightest dark spinor $\chi_1$ into four-components spinor $\Psi_1$, one has
\begin{equation}
    \L_{\Psi_1\bPsi_1} = i\frac{y_{11}^{a_0}}{2\sqrt{2}}(s_{\theta_A}A + c_{\theta_A}a)\bPsi_1\gamma_5\Psi_1
\end{equation}
Together, these enable $s$-wave annihilation $\chi_1\chi_1\to a/A\to b\bar b$, which remains unsuppressed at low velocities. A representative diagram is shown in Fig.~\ref{fig:halo_annihilation_diagram}, and the explicit annihilation cross sections are given in Appendix~\ref{app:ann_xsec}.

\begin{figure}[htb!]
    \centering
    \includegraphics[width=0.6\linewidth]{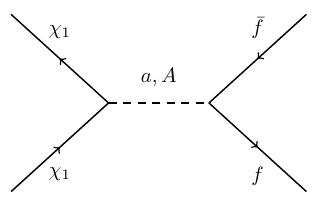}
    \caption{Broken-phase halo annihilation through CP-odd pseudoscalar exchange.}
    \label{fig:halo_annihilation_diagram}
\end{figure}

\section{Phenomenological Results}\label{sec:results}

We illustrate the phenomenology using three benchmark choices for the pseudoscalar masses $m_a$, $m_A$ and mixing angle $\theta_A$, summarized in Table~\ref{tab:benchmarks}. For all three benchmarks, we take $\tan\beta=18$, motivated by the current ATLAS type-II 2HDM heavy $H/A \to \tau\tau$ search limits around our benchmarks~\cite{ATLAS:2020zms}, and adopt the partial-alignment Yukawa relations
\begin{equation}
    y_R=y_L,\quad y_{RR}=-y_{LL}.
\end{equation}
We adopt partial alignment to simplify the scan. This is not a fundamental requirement on the model, but they can be motivated by a parity reflection
\begin{equation}\label{eq:parity_exchange}
    \chi_L\leftrightarrow\chi_R^c,\qquad
    \chi_M\to-\chi_M,\qquad
    \phi\to-\phi^*.
\end{equation}
Relaxing these relations provides additional freedom and further widen the allowed parameter regions. 

The remaining independent inputs are varied over the ranges summarized in Table~\ref{tab:scan_ranges}. We use a mixed scalar input basis involving physical masses, mixing angles, and selected potential parameters, with the remaining parameters reconstructed as described in Appendix~\ref{app:scalar_sector}. The scalar potential is required to satisfy perturbativity, tree-level unitarity, boundedness from below, and the neutral vacuum minimum conditions~\cite{Kanemura:2015ska,Goodsell:2018tti,Kannike:2012pe}. We also impose Higgs alignment~\cite{Heinemeyer:2021msz}, in order to obtain an SM-like CP even Higgs, and the relevant phenomenological constraints, including heavy scalar searches and DM direct detection.

\begin{table}[t]
    \centering
    \begin{tabular}{c c c c c}
        \hline
         & $m_a$ [TeV] & $m_A$ [TeV] & $\theta_A$ & $\tan\beta$ \\
        \hline
        BM1 & $0.9$ & $1.5$ & $0.6$ & $18$ \\
        BM2 & $0.9$ & $1.5$ & $0.4$ & $18$ \\
        BM3 & $1.1$ & $1.7$ & $0.6$ & $18$ \\
        \hline
    \end{tabular}
    \caption{}
    \label{tab:benchmarks}
\end{table}

\begin{table}[t]
    \centering
    \begin{tabular}{c c c}
        \hline
        Parameter & Range & Sampling \\
        \hline
        $m_H-m_A$
            & $[-0.5,\,0.5]~{\rm TeV}$ & linear \\
        $m_{H^\pm}-m_A$
            & $[-0.5,\,0.1]~{\rm TeV}$ & linear \\
        $|\theta_{S_{12}}|$
            & $0$--$0.3$ & linear \\
        $|\theta_{S_{23}}|$
            & $0$--$0.3$ & linear \\
        $v_\phi$
            & $0.1$--$5~{\rm TeV}$ & log \\
        $\lambda_1,\ |\lambda_{\phi1}|$
            & $10^{-5}$--$4\pi$ & log \\
        $M_\chi$
            & $0.3$--$5~{\rm TeV}$ & log \\
        $\Delta M\equiv M_D-M_\chi$
            & $0.05$--$5~{\rm TeV}$ & log \\
        $|y_R|$
            & $10^{-2}$--$\sqrt{8\pi}$ & log \\
        $|y_{RR}|$
            & $10^{-2}$--$\sqrt{8\pi}$ & log \\
        \hline
    \end{tabular}
    \caption{
        Scan ranges of the varying parameters.
    }
    \label{tab:scan_ranges}
\end{table}

\begin{figure*}[htb!]
    \centering
    \begin{subfigure}[t]{0.5\textwidth}
        \centering
        \includegraphics[width=\linewidth]
            {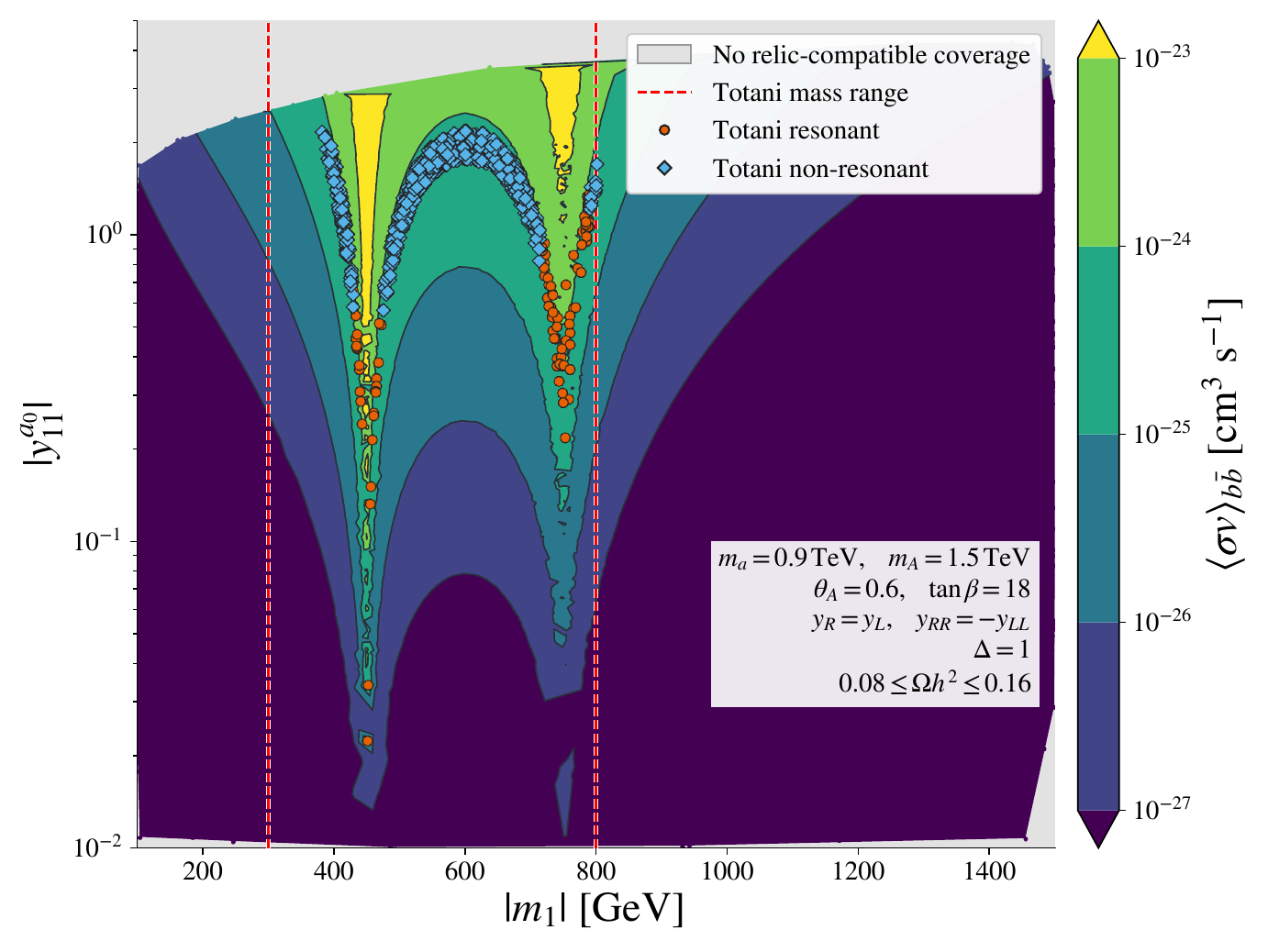}
        \caption{BM1}
        \label{fig:bm1_annihilation}
    \end{subfigure}\hfill
    \begin{subfigure}[t]{0.5\textwidth}
        \centering
        \includegraphics[width=\linewidth]
            {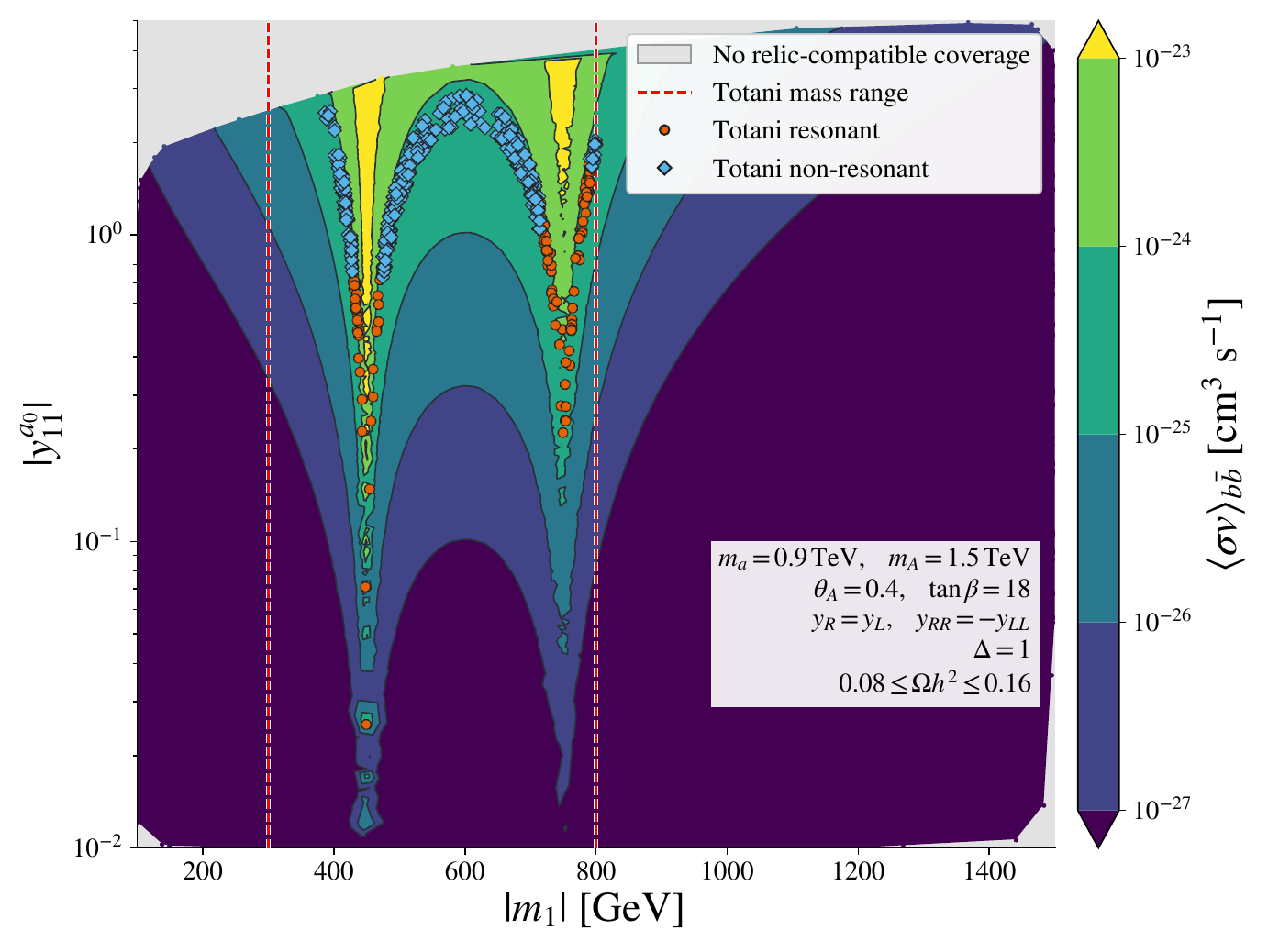}
        \caption{BM2}
        \label{fig:bm2_annihilation}
    \end{subfigure}\hfill
    \begin{subfigure}[t]{0.5\textwidth}
        \centering
        \includegraphics[width=\linewidth]
            {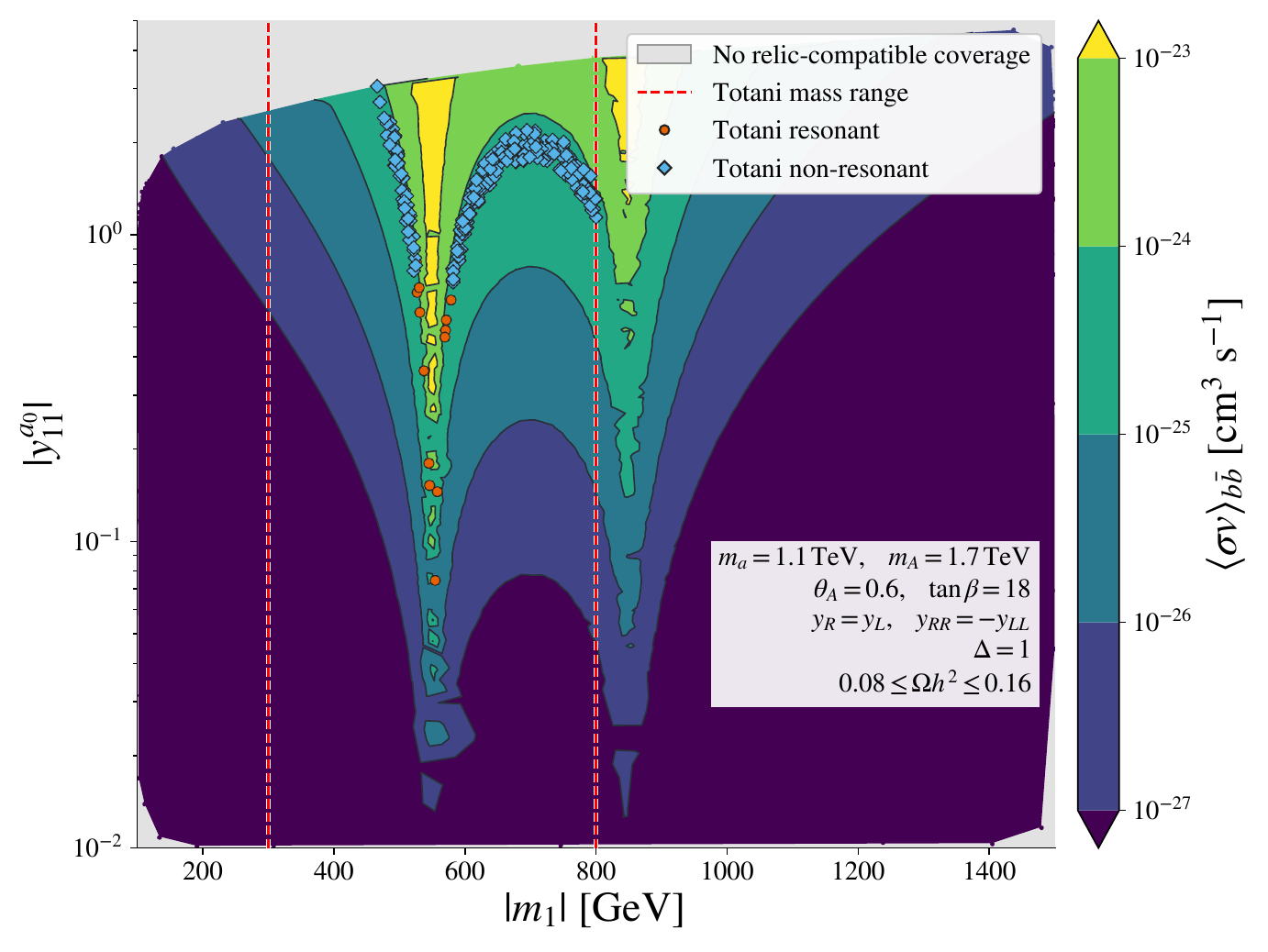}
        \caption{BM3}
        \label{fig:bm3_annihilation}
    \end{subfigure}
    \caption{
        Present day annihilation rate $\braket{\sigma v}_{b\bar b}$ in the $(|m_1|,|y_{11}^{a_0}|)$ plane for the three benchmarks in Table~\ref{tab:benchmarks}. Gray denotes regions without relic-compatible scan coverage. Orange circles and blue diamonds additionally reproduce the GHE target through resonant and non-resonant annihilation, respectively.
    }
    \label{fig:benchmark_annihilation}
\end{figure*}

Fig.~\ref{fig:benchmark_annihilation} shows the results for the three benchmarks in the $(|m_1|,|y_{11}^{a_0}|)$ plane. The colored contours interpolate the present day annihilation rate over the relic compatible sample~\cite{Planck:2018vyg}. The gray regions have no relic compatible coverage in the scan. Points that additionally reproduce the GHE target are marked by orange circles or blue diamonds, according to whether they lie or not near a pseudoscalar resonance~\cite{Totani:2025fxx}. We classify a point as resonant if
\begin{equation}
    \left|\frac{m_{a, A}-2|m_1|}{m_1}\right| < 0.1,
\end{equation}
and as non-resonant otherwise. All displayed points satisfy the theoretical consistency conditions, the direct detection limits~\cite{LZ:2024zvo}, and the other phenomenological constraints imposed in the scan as described in Appendix~\ref{app:scalar_sector}, together with the pre-BBN decay requirement in Eq.~\eqref{eq:BBN_decay}.

For BM1, the two resonance regions appear near $|m_1|=m_a/2\simeq450~{\rm GeV}$ and $|m_1|=m_A/2\simeq750~{\rm GeV}$. Close to either resonant pole, the required large annihilation rate can be reached with a relatively small DM Yukawa coupling $|y_{11}^{a_0}|$. Away from the poles, $|y_{11}^{a_0}| \sim \mathcal{O}(1)$ still produce the required rate while remain in compatible with relic density. The blue points demonstrate that the separation between thermal FO and present day annihilation does not require resonant enhancement.

BM2 retains the BM1 masses but reduces $\theta_A$ from $0.6$ to $0.4$. The resonance positions remain unchanged, while the weaker effective coupling of the pseudoscalars to SM fermions shifts the non-resonant GHE solutions toward larger $|y_{11}^{a_0}|$. BM3 instead increases the mediator masses. The corresponding poles move to $|m_1|\simeq550$ and $850~{\rm GeV}$, placing the heavier resonance above the GHE compatible mass range. Non-resonant solutions nevertheless remain in both benchmarks.

\begin{figure}[htbp]
    \centering
    \begin{subfigure}[b]{0.48\textwidth}
        \centering
        \includegraphics[width=\textwidth]{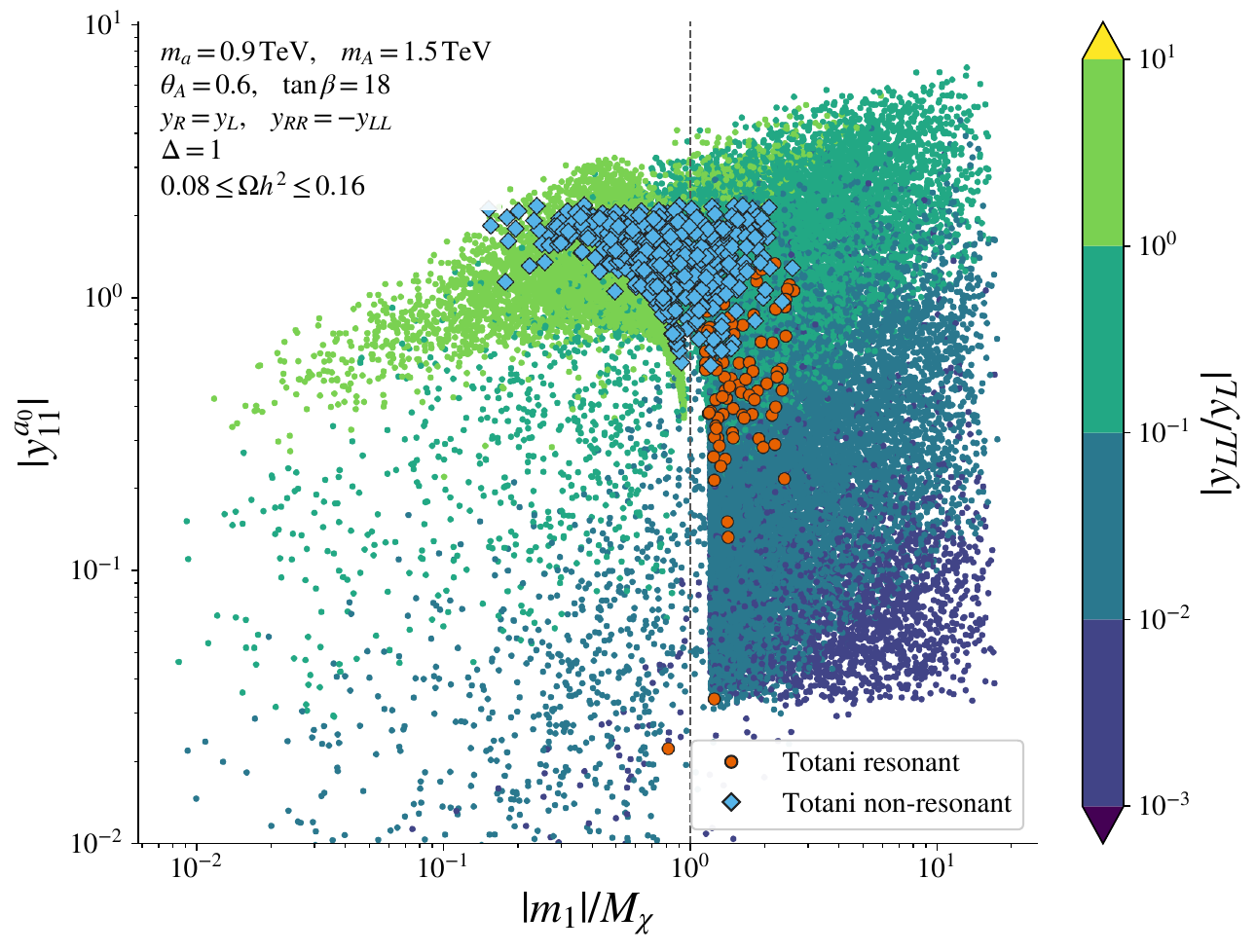}
        \caption{$\Delta = 1$}
        \label{fig:bm1_hierarchy_Delta1}
    \end{subfigure}
    \hfill 
    \begin{subfigure}[b]{0.48\textwidth}
        \centering
        \includegraphics[width=\textwidth]{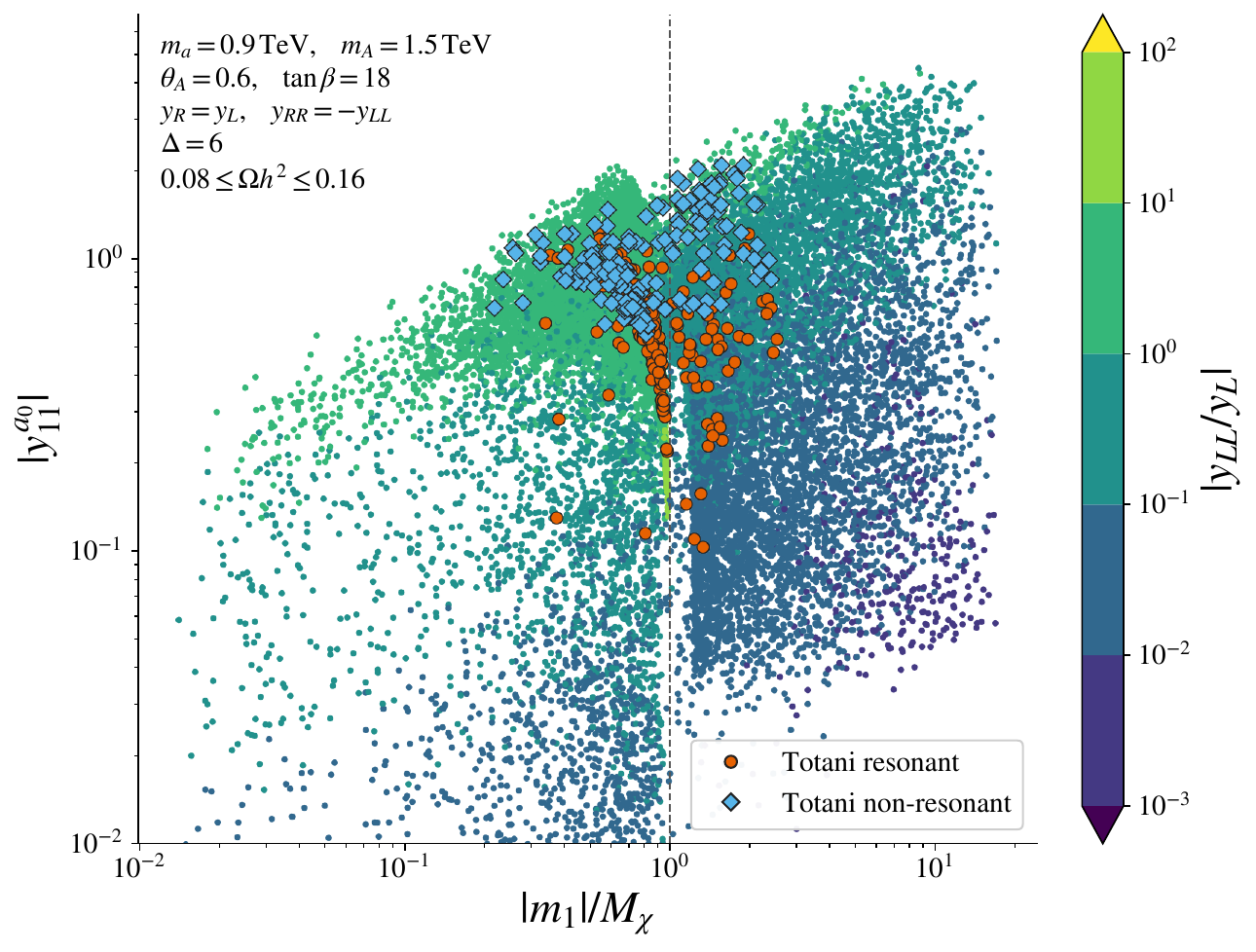}
        \caption{$\Delta = 6$}
        \label{fig:bm1_hierarchy_delta6}
    \end{subfigure}
    
    \caption{BM1 results in the $(|m_1|/M_\chi,|y_{11}^{a_0}|)$ plane. The color indicates $|y_{RR}/y_L|$, and the vertical dashed line marks $|m_1|=M_\chi$. Orange circles and blue diamonds additionally reproduce the GHE target through resonant and non-resonant annihilation, respectively.}
    \label{fig:bm1_hierarchy}
\end{figure}

The phase transition also changes the mass and composition of the surviving DM. Fig.~\ref{fig:bm1_hierarchy} displays the BM1 sample in the $(|m_1|/M_\chi,|y_{11}^{a_0}|)$ plane under different dilution assumptions, with the color indicating relative strengths of $|y_{LL}/y_L|$. The ratio $|m_1|/M_\chi$ compares the present DM mass with the mass of the state that underwent FO in early universe. GHE compatible points occur on both sides of unity, showing that the surviving relic can be either lighter or heavier than the original $\chi_M$. Generally, the surviving relic is lighter than the freeze-out $\chi_M$ when the ratio $|y_{LL}/y_L|$ is large and vice versa. The couplings $y_{LL}$ and $y_{RR}$ do not enter the leading symmetric-phase FO cross section, but they contribute to the broken-phase mass matrix and pseudoscalar couplings. Their variation therefore provides additional freedom in the present spectrum and annihilation rate. For both dilution benchmarks shown in Fig.~\ref{fig:bm1_hierarchy}, we find solutions that reproduce the observed relic abundance and the GHE target, including solutions away from the pseudoscalar resonances. These results show that our main conclusions remain robust even when there is substantial dilution from the phase transition.

\section{Discussion}\label{sec:discussion}

\subsection{Direct detection}

\begin{figure*}[htb!]
    \centering
    \captionsetup[subfigure]{labelformat=parens,labelsep=space,justification=centering}
    \begin{subfigure}[t]{0.32\textwidth}
        \centering
        \includegraphics[width=\linewidth]{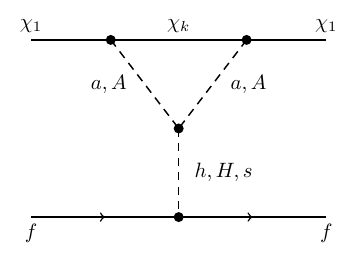}
        \caption{Triangle}
        \label{fig:si_loop_triangle}
    \end{subfigure}\hfill
    \begin{subfigure}[t]{0.32\textwidth}
        \centering
        \includegraphics[width=\linewidth]{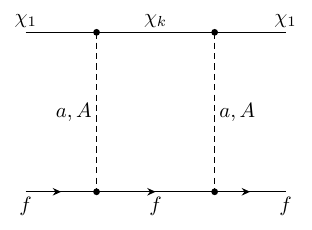}
        \caption{Box}
        \label{fig:si_loop_box}
    \end{subfigure}\hfill
    \begin{subfigure}[t]{0.32\textwidth}
        \centering
        \includegraphics[width=\linewidth]{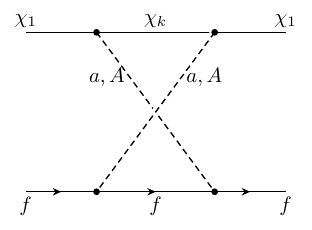}
        \caption{Crossed box}
        \label{fig:si_loop_crossed_box}
    \end{subfigure}
    \caption{One-loop pseudoscalar contributions to spin-independent scattering $\chi_1 f\to\chi_1 f$, where $f$ is an SM fermion.}
    \label{fig:si_loop_diagrams}
\end{figure*}

\begin{figure}[htb!]
    \centering
    \includegraphics[width=\linewidth]{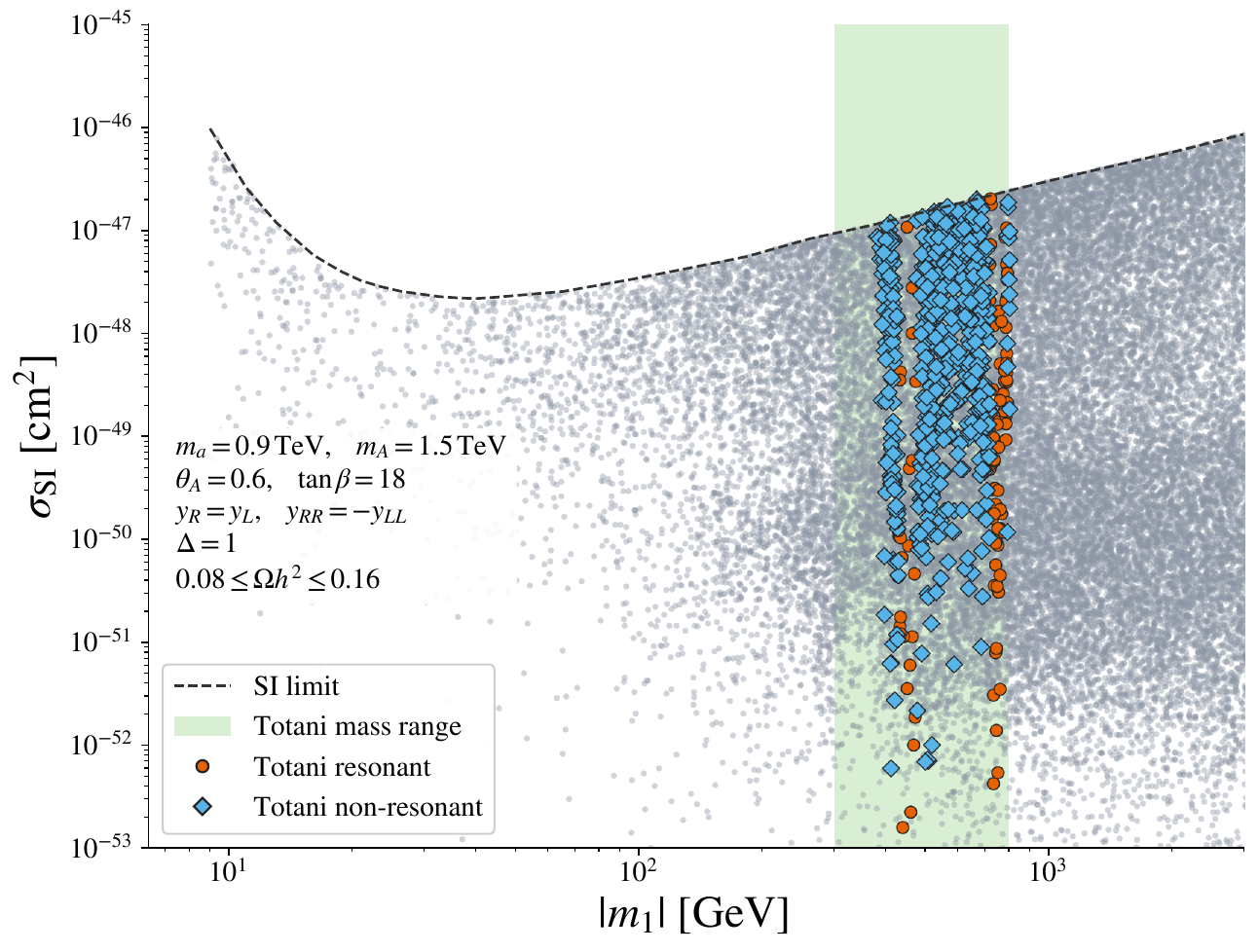}
    \caption{
        Spin-independent $\sigma_{\rm SI}$ vs. $|m_1|$. Orange circles and blue diamonds reproduce the GHE target through resonant and non-resonant annihilation, respectively. The green band indicates the GHE implied mass range. The dashed curve denotes the SI DD upper limit.
    }
    \label{fig:relic_direct_detection}
\end{figure}

In the non-relativistic limit relevant to direct detection, a CP-even scalar exchange gives an unsuppressed spin-independent response~\cite{Anand:2013yka},
\begin{equation}
    (\bar\chi\chi)(\bar N N)
    \;\to\;
    1,
\end{equation}
which goes to a constant regardless of the momentum transfer $\mathbf q$. The corresponding cross section is approximately
\begin{equation}\label{eq:tree_SI}
    \sigma_{\chi N}^{\rm SI, tree}
        \simeq
        \frac{\mu_{\chi N}^2m_N^2}{2\pi}
        \frac{|y_{11}^{s_0}|^2}{\vew^2}
        \left|
            \sum_{S\in\{H,h,s\}}
            \frac{R_{S3}}{m_S^2}f_N^{S,\rm SI}
        \right|^2,
\end{equation}
where $R_{Si}$ denotes the CP-even rotation from $(h_1,h_2,s_0)$ to the physical states $(H, h, s)$ (Appendix~\ref{app:scalar_sector}), and
\begin{equation}
    f_N^{S,\rm SI}
        =\delta_u^{\rm SI}\frac{R_{S2}}{\sin\beta}
        +(\delta_d^{\rm SI}+\delta_s^{\rm SI})
         \frac{R_{S1}}{\cos\beta}.
\end{equation}
For the neutron~\cite{Abe:2018emu},
\begin{equation}
    (\delta_u^{\rm SI},\delta_d^{\rm SI},\delta_s^{\rm SI})
        =(0.1469,\,0.0952,\,0.0447).
\end{equation}
In our model, small CP-even mixing suppresses this contribution without requiring the CP-odd mixing responsible for halo annihilation to be small.

By contrast, in the non-relativistic limit, a pseudoscalar exchange gives~\cite{Anand:2013yka}
\begin{equation}
    (\bar\chi i\gamma^5\chi)(\bar N i\gamma^5N)
    \to 
    \frac{
        (\mathbf S_\chi\cdot\mathbf q)(\mathbf S_N\cdot\mathbf q)
    }{
        m_\chi m_N
    } ,
\end{equation}
where $\mathbf S_\chi$ and $\mathbf S_N$ are the DM and nucleon spin operators, and $m_N$ is the nucleon mass. The elastic-scattering rate from CP-odd pseudoscalar exchange is therefore both spin-dependent and momentum suppressed. The corresponding cross section is approximately
\begin{equation}
    \sigma_{\chi N}^{\rm SD}
        \simeq
        \frac{\mu_{\chi N}^2 q^4}{32\pi m_1^2}
        \frac{
            |y_{11}^{a_0}|^2s_{\theta_A}^2c_{\theta_A}^2 (f_N^{\rm SD})^2
        }{
            \vew^2
        }
        \left|
            \frac{1}{m_a^2}
            -
            \frac{1}{m_A^2}
        \right|^2,
\end{equation}
where $\mu_{\chi N} = |m_1|m_N/(|m_1|+m_N)$ is the DM--nucleon reduced mass, and $f_N^{\rm SD}\sim \order{1}$--$\order{10}$ is the form factor encoding the nucleon response, including the $\tan\beta$ dependence. Since $|m_1| \sim 300$--$800$ GeV, $\mu_{\chi N}\sim m_N\simeq 1$ GeV. For characteristic $q\simeq100~{\rm MeV}$, $|y_{11}^{a_0}|\simeq1$, $\sin\theta_A\simeq0.5$, and representative masses $m_a \simeq 1$ TeV and $m_A \simeq 1.5$ TeV,
\begin{equation}
    \sigma_{\chi n}^{\rm SD}
        \sim 10^{-58} \text{--} 10^{-55}~{\rm cm^2},
\end{equation}
which is many orders of magnitude below current SD DD sensitivity $\sim\order{10^{-42}}~{\rm cm^2}$ for this DM mass range~\cite{LZ:2024zvo}.

The pseudoscalar portal, however, also induces SI scattering through the loop diagrams in Fig.~\ref{fig:si_loop_diagrams}~\cite{Abe:2018emu}. For the characteristic parameters considered here, the leading contribution comes from the $haa$ triangle. It benefits from the light Higgs propagator and the unsuppressed $\chi_1\chi_1a$ coupling, whereas the box diagrams require two mixing suppressed SM-pseudoscalar Yukawa vertices. Roughly
\begin{equation}
\begin{aligned}
    \sigma_{\chi N}^{\rm SI,loop}
        \sim{}& 10^{-51}~{\rm cm^2}\,
        |y_{11}^{a_0}|^4 c_{\theta_A}^4
        \left|\frac{\kappa_{haa}}{100~{\rm GeV}}\right|^2
        \\
        &\times
        \left(\frac{m_1}{500~{\rm GeV}}\right)^2
        \left(\frac{1~{\rm TeV}}{m_a}\right)^4,
\end{aligned}
\end{equation}
where $\kappa_{haa}$ is the trilinear scalar coupling. Including the other loop contributions does not change this estimate by orders of magnitude for the characteristic parameters considered here. Fig.~\ref{fig:relic_direct_detection} compares the numerically evaluated SI direct detection cross sections of our scanned points with the experimental upper limit. These include tree-level scalar exchange and loop contributions, with all allowed internal states included in the loop calculation.

\subsection{Indirect Detection}

Our mechanism naturally accommodates a present day annihilation cross section more than an order of magnitude larger than the thermal FO rate. The GHE-inferred cross section reported by Totani~\cite{Totani:2025fxx} is also in tension with dwarf spheroidal limits, although this discrepancy is considerably smaller than that with the canonical thermal FO rate and depends on astrophysical assumptions. Ref.~\cite{Stenhouse:2026nve} finds that, after accounting for uncertainties in Galactic diffuse-emission modeling and the MW and dwarf $J$-factors, the GHE inferred cross section can lie only around a factor of $2$ above the dwarf upper limit. Moreover, Ref.~\cite{Circiello:2026inp} points out that neglecting MW halo substructure, as in Totani's analysis, can overestimate the inferred GHE annihilation cross section. A further reduction of order two from this effect would suffice to close the residual discrepancy\footnote{Both Totani~\cite{Totani:2025fxx} and Ref.~\cite{Stenhouse:2026nve} assume a smooth NFW halo profile.}. Such a modest systematic variation leaves the main conclusions of our mechanism unchanged.

Alternatively, the dwarf tension can be addressed through the dSphobic mechanism of Ref.~\cite{Berlin:2025fwx}. Imposing the parity reflection $\chi_L\leftrightarrow\chi_R^c$, $\chi_M\to\pm\chi_M$, and $\phi\leftrightarrow\phi^*$ result in $y_R=\pm y_L$ and $y_{LL}=y_{RR}$\footnote{This is different from the alignment we adopted in Eq.~\eqref{eq:parity_exchange}}. This parity choice eliminates the diagonal pseudoscalar couplings. Instead, the two lightest broken-phase states, $\chi_1$ and $\chi_2$, retain an off-diagonal coupling, allowing $s$-wave coannihilation $\chi_1\chi_2\xrightarrow[]{a,A}f\bar f$. A tiny splitting $\delta_\chi\sim100~{\rm keV}$ can arise near a crossing of the aligned broken-phase mass eigenvalues. Its radiative stability would require additional approximate flavor symmetries, with the splitting generated by a small symmetry-breaking term. The tiny splitting closes SM charged fermion decay channels, while the $\chi_2\to\chi_1\gamma\gamma$ are strongly heavy-mediator loop suppressed, allowing $\chi_2$ to be cosmologically long-lived~\cite{Berlin:2025fwx}. One further introduces a light dark scalar $\varphi$ with only the off-diagonal coupling $y_\varphi\varphi\bar\chi_1\chi_2$, whose mass is in $\delta_\chi<m_\varphi\ll|m_{1,2}|$, to allow efficient conversion $\chi_2\chi_2\leftrightarrow\chi_1\chi_1$. Cosmological downscattering after the $\phi$ transition depletes the $\chi_2$ population once temperature drops below $\delta_\chi$. After structure formation, upscattering can replenish $\chi_2$ in the MW halo while remaining kinematically suppressed in dwarfs. The resulting population difference can suppress coannihilation in dwarfs while retaining a sizable MW signal. A quantitative treatment of this extension's cosmological and halo evolution is beyond the scope of this work.

\section{Conclusions}\label{sec:conclusions}

We have presented a phase transition portal scenario in which thermal DM can have an enhanced present day annihilation rate while remaining compatible with stringent DD bounds. Our concrete realization extends a type-II 2HDM by a complex scalar singlet, a Majorana fermion, and a heavier Dirac fermion. Before the singlet transition, the Majorana state freezes out through $p$-wave annihilation into singlet scalars. A delayed transition changes the dark fermion masses and mixing, opening $s$-wave annihilation of the surviving state through the mixed pseudoscalars $a$ and $A$. The interactions governing present day annihilation therefore differ from those that established the thermal abundance.

Using the reported Galactic Halo Excess~\cite{Totani:2025fxx} as an illustrative application, the scans exhibit relic compatible solutions with a present day annihilation rate more than an order of magnitude above the canonical thermal value without tuning in parameters.

The pseudoscalar interaction responsible for $s$-wave annihilation gives momentum-suppressed spin-dependent nuclear scattering. Small CP-even mixing suppresses the tree-level spin-independent response, and viable solutions remain after including the loop-induced contributions. Dwarf spheroidal limits present a separate, smaller tension for the GHE interpretation. Astrophysical uncertainties and MW halo substructure can remove this tension~\cite{Stenhouse:2026nve,Circiello:2026inp}, while the dSphobic extension discussed in Ref.~\cite{Berlin:2025fwx} provides an alternative particle-physics explanation through different dark-state populations in the MW and dwarf galaxies.

The mechanism is not specific to the GHE reported in Ref.~\cite{Totani:2025fxx} or to the chosen 2HDM implementation. An ALP-like portal with the same relevant pseudoscalar interactions can realize the same qualitative separation between FO and present day annihilation. A phase transition after FO thus provides a framework in which both the DM spectrum and its dominant annihilation process evolve, allowing a thermal relic to produce a large indirect signal with a suppressed nuclear recoil response.

\section*{Acknowledgments}
We thank M. Drees for useful discussions. The work of CW at the University of Chicago has been supported by the DOE grant DE-SC0013642. C.W. would like to thank the Aspen Center for Physics, which is supported by National Science Foundation grant No. PHY-1607611, where part of this work has been done. The work of PH is supported by the National Science
Foundation under grant number PHY-2412875. PH is grateful for the hospitality of Perimeter Institute where part of this work was carried out. Research at Perimeter Institute is supported in part by the Government of Canada through the Department of Innovation, Science and Economic Development and by the Province of Ontario through the Ministry of Colleges and Universities. This work was supported by a grant from the Simons Foundation (1034867, Dittrich). KZ gratefully acknowledges support from the Avenir Foundation.

\appendix
\section{Annihilation Cross Sections}\label{app:ann_xsec}

\subsection{Symmetric-phase $p$-wave annihilation}

The process $\chi_M\chi_M\to\phi\phi^*$ proceeds through $t$- and $u$-channel exchange of $\chi_D$. Neglecting the final scalar masses and retaining the leading nonrelativistic contribution, the thermal average is
\begin{equation}
\begin{aligned}
    &\langle{\sigma v_{rel}}\rangle_{y_L^4 + y_R^4}
        = \frac{(y_L^4 + y_R^4)}{32\pi xM_\chi^2} \left[ \frac{5}{(1 + r_D^2)^2} - \frac{4}{(1 + r_D^2)^3} + \frac{4}{(1 + r_D^2)^4} \right], \\
    &\langle{\sigma v_{rel}}\rangle_{y_L^2y_R^2}
        = \frac{y_L^2y_R^2}{16\pi xM_\chi^2} \\ 
        &\qquad \times \left[ \frac{6}{1+ r_D^2} - \frac{5}{(1 + r_D^2)^2} - \frac{4}{(1 + r_D^2)^3} + \frac{4}{(1 + r_D^2)^4}\right], \\
    &\langle{\sigma v_{rel}}\rangle_{(y_L^2 + y_R^2)y_Ly_R}
        = \frac{(y_L^2 + y_R^2)y_Ly_R r_D}{8\pi xM_\chi^2}\left[ \frac{3}{(1 + r_D^2)^2} - \frac{2}{(1 + r_D^2)^3} \right],
\end{aligned}
\end{equation}
where $r_D \equiv M_D / M_\chi$ and $x \equiv M_\chi/T$. We have used $\langle v_{\rm rel}^2\rangle=6/x$ for a Maxwell--Boltzmann distribution~\cite{Gondolo:1990dk}. Then,
\begin{equation}
    \langle{\sigma v_{rel}}\rangle 
        = \langle{\sigma v_{rel}}\rangle_{y_L^4 + y_R^4} + \langle{\sigma v_{rel}}\rangle_{y_L^2y_R^2} + \langle{\sigma v_{rel}}\rangle_{(y_L^2 + y_R^2)y_Ly_R}.
\end{equation}

\subsection{Present day $s$-wave annihilation}

The leading $s$-wave annihilation cross section for $\chi_1\chi_1\to b\bar b$ through $a$ and $A$ pseudoscalar exchange, away from the poles, is
\begin{equation}
\begin{split}
    \braket{\sigma v}_{b\bar b} 
        &= \frac{(1 - r_b^2)^{1/2}N_c s_{\theta_A}^2c_{\theta_A}^2(y_{11}^{a_0})^2m_b^2\tan^2\beta}{4\pi |m_1|^2\vew^2} \\
            &\qquad \times \left( \frac{1}{4 - r_a^2} - \frac{1}{4 - r_A^2} \right)^2
\end{split}
\end{equation}
where $r_X \equiv |m_X /m_1|$, and $N_c=3$ is the color factor.

\section{Scalar Lagrangian Parameters}\label{app:scalar_sector}

\begin{equation}\label{eq:field_expansion}
    H_i = 
    \begin{pmatrix}
        H^+_i \\
        \frac{1}{\sqrt{2}}(v_i + h_i + i\rho_i)
    \end{pmatrix}, \quad 
    \phi = \frac{1}{\sqrt{2}}(v_\phi + s_0 + ia_0)
\end{equation}
where $v^2 = v^2_1 + v_2^2$ and $\tan_\beta = v_2 / v_1$. Let
\begin{equation}
    m_{12,eff}^2 \equiv m_{12}^2 - \frac{\kappa_{12} v_\phi}{\sqrt{2}}.
\end{equation}

The tadpole conditions are
\begin{gather}
    m_{11}^2 - m_{12,eff}^2\tan\beta
        = - \frac{\lambda_1}{2}v_1^2 - \frac{\lambda_{3}+\lambda_4}{2}v_2^2 - \frac{\lambda_{\phi1}}{2}v_\phi^2, \label{eq:tadpole_m11} \\
    m_{22}^2 - m_{12,eff}^2 \cot\beta
        = - \frac{\lambda_2}{2}v_2^2 - \frac{\lambda_{3} + \lambda_4}{2}v_1^2 - \frac{\lambda_{\phi2}}{2}v_\phi^2, \\
    m_\phi^2
        = - \frac{\lambda_\phi}{4}v_\phi^2 - \frac{\mu_\phi}{2\sqrt{2}}v_\phi - \frac{\lambda_{\phi1}}{2}v_1^2 - \frac{\lambda_{\phi2}}{2}v_2^2 - \frac{\kappa_{12}}{\sqrt{2}}\frac{v_1v_2}{v_\phi} \label{eq:tadpole_mphi}.
\end{gather}

After imposing the tadpole conditions, the charged Higgs boson mass is
\begin{equation}
    m^2_{H^\pm} 
        = \Tr M^2_{H^\pm} 
        = \frac{m^2_{12,eff}}{s_\beta c_\beta} - \frac{\lambda_4}{2}\vew^2 .
\end{equation}
Before mixing, the CP-odd neutral Higgs boson mass is
\begin{equation}
    m_{A_0}^2 
        = \frac{m_{12,eff}^2}{s_\beta c_\beta}.
\end{equation}
and
\begin{equation}
    m^2_{a_0}
        = -\frac{3\mu_\phi v_\phi}{2\sqrt{2}} - \frac{\kappa_{12}}{\sqrt{2}}\frac{v_1v_2}{v_\phi},
\end{equation}
\begin{equation}
    M^2_{(A_0,a_0)}
        = 
        \begin{pmatrix}
            m^2_{A_0} & -\kappa_{12} \vew/\sqrt{2} \\
            -\kappa_{12} \vew /\sqrt{2} & m_{a_0}^2
        \end{pmatrix}.
\end{equation}
A rotation $(A_0, a_0) \to (A, a)$ by mixing angle
\begin{equation}
    \tan 2\theta_A 
        = \frac{2M^2_{A_0a_0}}{M_{A_0A_0}^2 - M_{a_0a_0}^2}
        = \frac{-\sqrt 2\kappa_{12}\vew}{m^2_{A_0} - m_{a_0}^2}
\end{equation}
diagonalizes the matrix. The eigenvalues after mixing are
\begin{gather}
    m^2_a
        = \frac{1}{2}\left[ m^2_{A_0} + m_{a_0}^2 - \sqrt{(m^2_{A_0} - m_{a_0}^2)^2 + 2\kappa_{12}^2\vew^2} \right], \\
    m^2_A
        = \frac{1}{2}\left[ m^2_{A_0} + m_{a_0}^2 + \sqrt{(m^2_{A_0} - m_{a_0}^2)^2 + 2\kappa_{12}^2\vew^2} \right],
\end{gather}
and the pre- and post-mixing states are related via
\begin{gather}
    A_0 = c_{\theta_A} A - s_{\theta_A}a, \\
    a_0 = s_{\theta_A} A + c_{\theta_A}a, \\
    m^2_{A_0} = m^2_Ac^2_{\theta_A} + m^2_a s^2_{\theta_A}, \\
    m^2_{a_0} = m^2_As^2_{\theta_A} + m^2_a c^2_{\theta_A}, \\
    m^2_{A_0a_0} = |m^2_A - m^2_a|s_{\theta_A}c_{\theta_{A}}.
\end{gather}

For the CP-even states, we define
\begin{equation}
    (H, h, s)^T = R_{\theta_S} (h_1, h_2, s_0)^T
\end{equation}
with~\cite{Heinemeyer:2021msz}
\begin{equation}
    R_{\theta_S}
        = 
        \begin{pmatrix}
            c_{12}c_{13} & s_{12}c_{13} & s_{13} \\
            -s_{12}c_{23} - c_{12}s_{23}s_{13} & c_{12}c_{23} - s_{12}s_{23}s_{13} & s_{23}c_{13} \\
            s_{12}s_{23} - c_{12}c_{23}s_{13} & -c_{12}s_{23} - s_{12}c_{23}s_{13} & c_{23}c_{13}
        \end{pmatrix},
\end{equation} 
where $c_{ij}\equiv\cos\theta_{S_{ij}}$ and $s_{ij}\equiv\sin\theta_{S_{ij}}$. 

In the basis $(h_1,h_2,s_0)$, after imposing the tadpole conditions, its entries are
\begin{gather}
    M_{11}^2 
        = m^2_{12,eff}\tan\beta + \lambda_1v_1^2, \label{eq:M11}\\
    M_{22}^2 
        = m^2_{12,eff}\cot\beta + \lambda_2v_2^2, \\
    M_{12}^2 
        = -m^2_{12,eff} + (\lambda_{3} + \lambda_4)v_1v_2, \\
    M_{13}^2 
        = \lambda_{\phi 1} v_\phi v_1 + \frac{\kappa_{12}}{\sqrt{2}}v_2, \\
    M_{23}^2 
        = \lambda_{\phi 2}v_\phi v_2  + \frac{\kappa_{12}}{\sqrt{2}}v_1, \\
    M_{33}^2 
        = \frac{\lambda_\phi}{2}v_\phi^2 + \frac{\mu_\phi}{2\sqrt{2}}v_\phi - \frac{\kappa_{12}}{\sqrt{2}}\frac{v_1v_2}{v_\phi}\label{eq:M33}
\end{gather}
Equivalently, it can be reconstructed from the physical masses as
\begin{equation}\label{eq:scalar_matrix_elements}
    M^2_{S_{ij}} 
        = \sum^3_{k = 1} (R_{\theta_S})_{ki}(R_{\theta_S})_{kj} m^2_{k}.
\end{equation}

\subsection{Parameter Reconstruction}

For the numerical scan, we use the physical values
\[
    m_h, m_H, m_A, m_{H^\pm}, m_a, \theta_{S_{12}}, \theta_{S_{23}}, \theta_{A}, \tan\beta, \vew, v_\phi,
\]
and the Lagrangian parameters
\[
    \lambda_1, \lambda_{\phi1}
\]
as the scan inputs. All other parameters can be inverted from them.

The CP-odd and charged sectors determine
\begin{gather}
    \kappa_{12}
        =-\frac{m_A^2-m_a^2}{\sqrt{2}\vew}
          \sin2\theta_A,
    \\
    m_{12,{\rm eff}}^2
        =s_\beta c_\beta
          \left(m_A^2c_{\theta_A}^2
               +m_a^2s_{\theta_A}^2\right),
    \\
    \mu_\phi
        =-\frac{2\sqrt{2}}{3v_\phi}
          \left(
              m_A^2s_{\theta_A}^2
              +m_a^2c_{\theta_A}^2
              +\frac{\kappa_{12}v_1v_2}{\sqrt{2}v_\phi}
          \right),
    \\
    \lambda_4
        =\frac{2m_{12,{\rm eff}}^2}{v_1v_2}
          -\frac{2m_{H^\pm}^2}{\vew^2}.
\end{gather}

The remaining physical parameters, $m_s^2$ and $\theta_{S_{13}}$, follow from
\begin{gather}
    \sum_{k=H,h,s}(R_{\theta_S})_{k1}^2m_k^2
        =m_{12,{\rm eff}}^2\tan\beta+\lambda_1v_1^2, \\
    \sum_{k=H,h,s}
        (R_{\theta_S})_{k1}(R_{\theta_S})_{k3}m_k^2
        =\lambda_{\phi1}v_\phi v_1
          +\frac{\kappa_{12}}{\sqrt{2}}v_2.
\end{gather}
We require real solutions with $m_s^2>0$. Using Eq.~\eqref{eq:scalar_matrix_elements}, the remaining quartic couplings are
\begin{gather}
    \lambda_2 
        = \frac{M^2_{22} - m^2_{12,eff}\cot\beta}{v_2^2}, \\
    \lambda_3
        = \frac{M^2_{12} + m^2_{12,eff}}{v_1v_2} - \lambda_4, \\
    \lambda_{\phi2} 
        = \frac{M_{23}^2 - \frac{\kappa_{12}}{\sqrt{2}}v_1}{v_\phi v_2}, \\
    \lambda_\phi
        = \frac{2M_{33}^2 - \frac{\mu_\phi}{\sqrt{2}}v_\phi + \sqrt{2}\kappa_{12}\frac{v_1v_2}{v_\phi}}{v_\phi^2}.
\end{gather}
The $m_{11}^2$, $m_{22}^2$, and $m^2_\phi$ are then reconstructed with the tadpole conditions Eq.~\eqref{eq:tadpole_m11}--\eqref{eq:tadpole_mphi}.

The scalar trilinear couplings $\kappa_{ijk}$ can then be obtained directly from the potential in the physical basis as
\begin{equation}
    -i\kappa_{ijk}
        =-i \frac{\partial^3V}{\partial\Phi_i\,\partial\Phi_j\,\partial\Phi_k}
        \bigg|_{\rm vac}, \quad \Phi_i\in\{h,H,s,a,A\}.
\end{equation}

\subsection{Theoretical Consistency}

We require the reconstructed scalar potential to satisfy perturbativity, boundedness from below~(BFB), and tree-level unitarity. Also, the chosen vacuum must be locally stable and be the global minimum in the field space.

\subsubsection{Perturbativity}

We impose the perturbativity condition
\begin{equation}
    |\lambda|<4\pi,
    \quad
    \lambda\in
    \{\lambda_{1,2,3,4},
      \lambda_\phi,\lambda_{\phi1},\lambda_{\phi2}\}.
\end{equation}

\subsubsection{Bounded from Below~(BFB)}

Stability at large field values is controlled by the quartic potential. Define
\[
    r_i \equiv H^\dagger_i H_i \geq 0, \quad |H_1^\dagger H_2| \equiv \sqrt{r_1 r_2x},\  0\leq x \leq 1, \quad r_\phi \equiv |\phi|^2.
\]
Then, the quartic potential can be expressed as
\begin{equation}
\begin{split}
    V_4 
        &= \frac{\lambda_1}{2}r_1^2 + \frac{\lambda_2}{2}r_2^2 + \lambda_3r_1 r_2 + \lambda_4 xr_1r_2 \\ 
        &\qquad + \frac{\lambda_\phi}{4}r_\phi^2 + \lambda_{\phi1}r_\phi r_1 + \lambda_{\phi2}r_\phi r_2 \\
        &\equiv \vec{r}^T A(x)\vec{r}
\end{split}
\end{equation}
for $\vec{r} \equiv (r_1, r_2, r_\phi)^T$ and where
\begin{equation}
    A(x)
        \equiv 
        \begin{pmatrix}
            \lambda_1 / 2 & (\lambda_3 + \lambda_4 x)/2 & \lambda_{\phi 1}/2 \\
            (\lambda_3 + \lambda_4 x)/2 & \lambda_2 / 2 & \lambda_{\phi2}/2 \\
            \lambda_{\phi 1}/2 & \lambda_{\phi 2}/2 & \lambda_\phi /4
        \end{pmatrix}.
\end{equation}
BFB is equivalent to requiring that $A(x)$ is copositive for every allowed $x$. For a $3 \times 3$ real symmetric matrix, the copositivity conditions are analytically given by requiring~\cite{Kannike:2012pe}
\begin{gather}
    a_{11}\ge0,\qquad a_{22}\ge0,\qquad a_{33}\ge0, \\
    a_{12}+\sqrt{a_{11}a_{22}}\ge0, \\
    a_{13}+\sqrt{a_{11}a_{33}}\ge0, \\
    a_{23}+\sqrt{a_{22}a_{33}}\ge0, \\
    \sqrt{a_{11}a_{22}a_{33}} + a_{12}\sqrt{a_{33}} + a_{13}\sqrt{a_{22}} + a_{23}\sqrt{a_{11}} \notag \\ + \sqrt{2\,(a_{12}+\sqrt{a_{11}a_{22}})(a_{13}+\sqrt{a_{11}a_{33}})(a_{23}+\sqrt{a_{22}a_{33}})} \ge 0.
\end{gather}
We consider the worst case of which the $\lambda_3 + \lambda_4x$ takes its minimum value
\[
    \lambda^{\rm worst}_{34} = \lambda_3 + \min (0, \lambda_4).
\]
Then, we can just test copositivity of 
\begin{equation}
    A
        \equiv 
        \begin{pmatrix}
            \lambda_1 / 2 & \lambda^{\rm worst}_{34}/2 & \lambda_{\phi 1}/2 \\
            \lambda^{\rm worst}_{34}/2 & \lambda_2 / 2 & \lambda_{\phi2}/2 \\
            \lambda_{\phi 1}/2 & \lambda_{\phi 2}/2 & \lambda_\phi /4
        \end{pmatrix}.
\end{equation}

\subsubsection{Tree-level Unitarity}

We impose tree-level perturbative unitarity using the high energy two-scalar scattering matrix constructed from the quartic potential, including the Goldstone modes, following the procedures in Refs.~\cite{Kanemura:2015ska,Goodsell:2018tti}. All eigenvalues $\Lambda_i$ of the scattering matrix in an orthonormal two particle basis are required to satisfy $|\Lambda_i|\leq8\pi$.

\subsubsection{Local and Global Minimum}

The tadpole conditions in Eqs.~\eqref{eq:tadpole_m11}--\eqref{eq:tadpole_mphi} make the chosen CP-even neutral vacuum $(v_1,v_2,v_\phi)$ stationary. Local stability requires positive squared masses in all physical scalar directions. We test the global-minimum condition by numerically minimizing the tree-level potential and discard points for which a deeper minimum is found.

\subsection{Phenomenological Constraints}\label{app:scalar_pheno}

The numerical scans are further tested against phenomenological constraints. The SM-like Higgs boson $h$'s couplings are rescaled as
\begin{equation}\label{eq:kappa_hff}
\begin{aligned}
    \kappa_{hVV}
        &=c_\beta R_{h1}+s_\beta R_{h2},
    \\
    \kappa_{hdd}=\kappa_{h\ell\ell}
        &=\frac{R_{h1}}{c_\beta},
    \quad
    \kappa_{huu}
        =\frac{R_{h2}}{s_\beta},
\end{aligned}
\end{equation}
where $V=W,Z$. We impose the alignment requirement~\cite{Heinemeyer:2021msz}
\begin{equation}
    |\kappa_{hVV}|>0.99,
    \quad
    0.8<|\kappa_{hdd}|<1.2.
\end{equation}

For the charged Higgs boson, we adopt the type-II constraint from $b\to s\gamma$~\cite{Baer:2023uwo,Misiak:2026sqy}
\begin{equation}
    m_{H^\pm}>670~{\rm GeV}.
\end{equation}
To close the exotic Higgs decay channels $h\to aa,ss$, we also require
\begin{equation}
    m_{s, a}>m_h / 2.
\end{equation}

Heavy neutral scalar searches in the $\tau^+\tau^-$ channel are checked against the ATLAS exclusion contour~\cite{ATLAS:2020zms}. The ATLAS benchmark is approximately in the alignment limit, where both $A$ and $H$ have down-type Yukawa couplings enhanced by $\tan\beta$ relative to the SM. In our model, the corresponding coupling modifiers include the additional scalar mixing as
\begin{equation}\label{eq:kappa_Phiff}
\begin{aligned}
    \kappa_{A\ell\ell}
        &=\tan\beta\,c_{\theta_A},
    &
    \kappa_{a\ell\ell}
        &=-\tan\beta\,s_{\theta_A},
    \\
    \kappa_{H\ell\ell}
        &=\frac{R_{H1}}{c_\beta},
    &
    \kappa_{s\ell\ell}
        &=\frac{R_{s1}}{c_\beta}.
\end{aligned}
\end{equation}
Hence, we approximate each heavy neutral constraint by
\begin{equation}
    |\kappa_{\Phi \ell\ell}|
        < \tan\beta_{\rm ATLAS}^{\rm lim}(m_\Phi), \quad \Phi\in\{A,a,H,s\}.
\end{equation}
Additional decay channels, particularly $\Phi\to\chi_i\chi_j$, reduce ${\rm BR}(\Phi\to\tau^+\tau^-)$ and further weaken the experimental constraint~\cite{Baer:2022qqr,Baer:2022smj,Baer:2023yxk,Baer:2025zqt,Baer:2025uzx}. We neglect this branching fraction dilution, making our approximate constraint more stringent where these channels are open.

\section{$\chi$ Decay}\label{app:chi_decay}

The heavier fermions $\chi_{2,3}$ decay into $\chi_1$ through channels involving the neutral scalars $\Phi\in\{h,H,s,a,A\}$. Write the mass basis interactions as
\begin{equation}
    \L_Y
        \supset -\frac{1}{2}\sum_{\Phi,i,j} g_{ij}^{\Phi}\Phi\chi_i\chi_j+\hc,
\end{equation}
where
\begin{equation}\label{eq:chi_coup}
\begin{aligned}
    g_{ij}^{S}
        &=\frac{R_{S3}}{\sqrt{2}}y_{ij}^{s_0},
        && S\in\{h,H,s\},
    \\
    g_{ij}^{a}
        &=\frac{i c_{\theta_A}}{\sqrt{2}}y_{ij}^{a_0},
    &\qquad
    g_{ij}^{A}
        &=\frac{i s_{\theta_A}}{\sqrt{2}}y_{ij}^{a_0}.
\end{aligned}
\end{equation}
The phases inherited from the fermion diagonalization are retained in these couplings.

For $|m_{j}|>|m_{i}|+m_\Phi$, the partial decay width is~\cite{Dreiner:2008tw}
\begin{equation}
\begin{aligned}
    \Gamma(\chi_j\to\chi_i\Phi)
    ={}&
    \frac{
        \lambda^{1/2}
        (|m_{j}|^2,|m_{i}|^2,m_\Phi^2)
    }{16\pi |m_{j}|^3}
    \\
    &\times
    \Bigl\{
        [\operatorname{Re}g_{ij}^{\Phi}]^2
        \bigl[(|m_{j}|+|m_{i}|)^2-m_\Phi^2\bigr]
        \\
        &\qquad
        +[\operatorname{Im}g_{ij}^{\Phi}]^2
        \bigl[(|m_{j}|-|m_{i}|)^2-m_\Phi^2\bigr]
    \Bigr\},
\end{aligned}
\label{eq:chi_two_body_decay}
\end{equation}
where $\lambda(x,y,z)=x^2+y^2+z^2-2xy-2xz-2yz$ is the K\"all\'en function.

Even when all $2$-body decays of $\chi_j$ are kinematically closed, the off-shell neutral scalars $\Phi$ can mediate $3$-body $\chi_{2, 3}\to\chi_1 f\bar f$. These channels are important in a compressed spectrum $m_f\ll\Delta_{ji}\ll|m_i|,m_\Phi$, where $\Delta_{ji}\equiv|m_j|-|m_i|$. The partial width is roughly~\cite{Djouadi:2001fa,Berlin:2025fwx}
\begin{equation}
\begin{aligned}
    \Gamma(\chi_j\to\chi_i f\bar f)
        &\simeq{}
        \frac{N_c^f\Delta_{ji}^5}{60\pi^3}
        \Biggl[
            (\operatorname{Re}C_S^f)^2
            +(\operatorname{Im}C_P^f)^2
            \\
            &+\frac{3\Delta_{ji}^2}
            {7(|m_j|+|m_i|)^2}
            \left\{
                (\operatorname{Im}C_S^f)^2
                +(\operatorname{Re}C_P^f)^2
            \right\}
    \Biggr],
\end{aligned}
\label{eq:chi_three_body_visible_estimate}
\end{equation}
where
\begin{equation}
    C_S^f
        =-\sum_{S=h,H,s}
          \frac{g_{ij}^Sg_{ff}^S}{m_S^2},
    \quad
    C_P^f
        =-\sum_{P=a,A}
          \frac{ig_{ij}^Pg_{ff}^P}{m_P^2},
\end{equation}
\begin{equation}
    g_{ff}^\Phi=(m_f/\vew)\kappa_{\Phi ff}
\end{equation}
with $\kappa_{\Phi ff}$ given in Eq.~\eqref{eq:kappa_hff} and Eq.~\eqref{eq:kappa_Phiff}. The three-body decay is only used when two-body decay channels are closed. 

\bibliography{ref}

\end{document}